\documentclass[12pt,letterpaper]{article}
\usepackage[utf8]{inputenc}
\usepackage{amsmath}
\usepackage{mathptmx, courier,textcomp}
\usepackage{amsfonts}
\usepackage{amssymb}
\usepackage{graphicx}
\usepackage{cite}
\usepackage{ulem}
\usepackage{caption}
\usepackage{epstopdf}
\usepackage[top=1in,bottom=1in,left=1in,right=1in]{geometry} 
\usepackage{float}
\usepackage{hyperref}
\usepackage{booktabs}
\usepackage[export]{adjustbox}
\usepackage{float}
\usepackage{multirow}
\usepackage{pifont}
\usepackage[ ]{authblk}
\usepackage{chemarrow}
\usepackage{comment}
\usepackage{makecell}
\UseRawInputEncoding
\hypersetup{
    colorlinks=true,
    linkcolor=blue,
    filecolor=magenta,      
    urlcolor=blue,
    pdftitle={Overleaf Example},
    pdfpagemode=FullScreen,
}

\begin{document} 

%
%
\title{Enhancing Electronic and Optical Properties of $\alpha$-Fe$_2$O$_3$ by Introducing B, Y, and Nb Dopants for Improved Photoelectrochemical Water Splitting} 
%
%

%
%
\author{Abdul Ahad Mamun}
\author[1,*]{Muhammad Anisuzzaman Talukder}
\affil{\small{Department of Electrical and Electronic Engineering\\

Bangladesh University of Engineering and Technology\\

Dhaka 1205, Bangladesh\\}}
\affil[ ]{\small{\it{$^*$anis@eee.buet.ac.bd}}}
%
%

\date{}
\maketitle

\newcommand{\OO}{O$_2$~}
\newcommand{\HH}{H$_2$~}
\sloppy
%
%
\begin{abstract}
Advanced theoretical investigations are crucial for understanding the structural growth mechanisms, optoelectronic properties, and photocatalytic activity of photoelectrodes for efficient photoelectrochemical water splitting. In this work, we conducted first-principles calculations aimed at designing $\alpha$-Fe$_2$O$_3$ photoelectrodes incorporating mono-dopants such as boron (B), yttrium (Y), and niobium (Nb), as well as co-dopants (B, Y) and (B, Nb) to enhance the performance of photoelectrochemical cells (PECs). We assessed the thermodynamic phase stability by calculating formation enthalpy ($E_f$) and examining material properties, including microstrain ($\mu_\epsilon$) and crystallite size ($D$). The mono-dopants, Y and Nb, and the co-dopants, (B, Y) and (B, Nb), exhibited negative $E_f$ values under the substitutional doping method, confirming their thermodynamic phase stability and suggesting their practical viability for experimental implementation. Notably, the values of $\mu_\epsilon$ and $D$ fell within the ranges observed experimentally for $\alpha$-Fe$_2$O$_3$, indicating their effectiveness in growth mechanisms. To gain a comprehensive understanding of the optoelectronic properties of doped $\alpha$-Fe$_2$O$_3$, we calculated the electronic band structure, density of states, atom's ionic charge, and optical absorption coefficient ($\alpha$). This analysis allowed us to examine the improvements in the electronic charge characteristics and photon-electron interactions. B-doped $\alpha$-Fe$_2$O$_3$ led to the formation of impurity bands, which were effectively mitigated by utilizing co-dopants (B, Y) and (B, Nb). The metal dopants, Y and Nb, significantly increased the charge carrier density, while the co-dopants, (B, Y) and (B, Nb), substantially enhanced light absorption in the visible spectrum. These improvements in the electronic and optical properties of $\alpha$-Fe$_2$O$_3$ indicate its potential for application in photocatalytic water splitting.
\end{abstract}
%
%

%
%

%
%
\section{Introduction}

Converting solar energy into renewable hydrogen (H$_2$) fuel through water (H$_2$O) splitting reactions in photoelectrochemical cells (PECs) is considered a promising third-generation energy conversion technology \cite{chiu2019photoelectrochemical,jia2016solar}. It generates zero carbon emissions and presents a favorable alternative to fossil fuels by enabling sustainable and efficient PEC performance \cite{chiu2019photoelectrochemical,li2023solar}. The U.S.~Department of Energy has established standard performance targets for PEC systems, requiring $\sim$10\% solar-to-hydrogen efficiency ($\eta_{\rm STH}$) and $\sim$1000 hours of durability, both of which critically depend on the photoelectrode \cite{li2013photoelectrochemical,thakur2020current}. Semiconductor materials serve as photoelectrodes by absorbing sunlight to generate electron-hole pairs, resulting in photoinduced current density ($J_{\rm ph}$) and potential ($V$) \cite{chiu2019photoelectrochemical,sivula2016semiconducting}. In PECs, the produced ($V$) must exceed the combined values of the standard water-splitting redox potential ($E_{\rm rx}^0$) and the losses associated with the PEC system to facilitate charge carrier transfer at the interfaces between the photoelectrode and the electrolyte for oxygen (OER) and hydrogen evolution reactions (HER) \cite{mamun2023effects,hemmerling2021design}. To meet the requirements of the PEC system, the photoelectrode should have a bandgap energy ($E_g$) in the range of 1.6 to 2.0 eV and achieve a minimum $J_{\rm ph}$ of 8.2 mAcm$^{-2}$ \cite{walter2011correction,che2021photocatalytic}. Furthermore, enhancing the kinetic rate of evolution reactions and ensuring the electrochemical stability of the photoelectrode are crucial for efficient H$_2$ production and longevity \cite{qureshi2024photoelectrochemical,mamun2024enhancing}. These advancements are key to unlocking the full potential of solar H$_2$ as a clean and renewable energy source for the future.

In recent decades, extensive research has explored various semiconductors as photoelectrodes, including compound semiconductors (CSs), transition metal oxides (TMOs), transition metal carbides (TMCs), metal nitrides (MNs), metal selenides (MSs), and metal chalcogenides (MCs) \cite{sivula2016semiconducting,gao2021transition,diez2022progress,hamdani2021recent,jian2020review,higashi2022design}. Among these, TMOs---chemical compounds comprising various transition metals and oxygen---represent a promising class of materials to use as photoelectrodes for advancing PEC systems due to technically and economically feasible synthesis processes \cite{lu2022photoelectrocatalytic,barba2021oxide}. TMOs offer several advantages, including the presence of oxygen vacancy defect sites, tunable $E_g$, and diverse morphological structures \cite{lee2019progress}. Additionally, they are low-cost, abundant in nature, and demonstrate exceptional photochemical stability in electrolytes under sunlight \cite{yao2018photoelectrocatalytic}. 

Despite these advantages, the widespread application of TMOs in PEC systems faces significant challenges. They inherently form polarons and comprise charge carrier quasi-particles, resulting in inefficient electron-hole pair separations \cite{wang2018metal}. Specifically, TMOs have a larger $E_g$, limiting their ability to absorb a significant portion of solar irradiance, leading to insufficient charge carrier generation and $J_{\rm ph}$ \cite{tan2022state,wang2018metal}. Furthermore, rapid charge carrier recombination within the bulk material and slower charge transfer kinetics at the interface between TMOs and electrolytes present critical barriers to large-scale H$_2$ production via photoelectrochemical water splitting \cite{yang2024advances}. Recently, researchers have focused on developing new methods and approaches to enhance the optoelectronic properties and photocatalytic activity of various TMOs, such as $\alpha$-Fe$_2$O$_3$, TiO$_2$, BiVO$_4$, WO$_3$, CuO, and CdWO$_4$. \cite{mali2015photoelectrochemical,zhou2020interstitial,singh2015tailoring}. 

$\alpha$-Fe$_2$O$_3$, commonly known as hematite, is regarded as one of the most effective photoelectrodes for PEC applications due to several key factors: the favorable oxidized states of Fe atom (Fe$^{2+}$ and Fe$^{3+}$), its substantial abundance in the Earth's crust, non-toxicity, and high electrochemical stability across a wide range of pH levels in electrolytes \cite{li2013photoelectrochemical,najaf2021recent}. Notably, the band edge position relative to the redox potential of water oxidation, along with a relatively suitable $E_g$ of $\sim$2.2 eV, makes $\alpha$-Fe$_2$O$_3$ a preferred material for photoanodes, enabling it to ideally utilize $\sim$40\% of solar energy for photoabsorption \cite{kay2006new}. If all photons with energy $>$2.2 eV produce electron-hole pairs with 100\% quantum yield efficiency, the $J_{\rm ph}$ and power conversion efficiency ($\eta_{\rm PCE}$) of $\alpha$-Fe$_2$O$_3$ could theoretically reach $\sim$12.6 mAcm$^{-2}$ and $\sim$12.9\% in photoelectrochemical water splitting \cite{murphy2006efficiency}. However, the use of pristine $\alpha$-Fe$_2$O$_3$ in the PEC system is limited due to its poor electrical conductivity ($<1$ cm$^2$V$^{-1}$s$^{-1}$), a rapid charge carrier recombination rate ($\sim$10 ps), a short diffusion length (2--4 nm), and the difficulty of exciting d-orbital electrons with only a small portion of visible light (up to $<$564 nm) \cite{franking2013facile,gardner1963electrical,balberg1978optical}. Additionally, insufficient charge carrier separation and a less favorable surface structure of pristine $\alpha$-Fe$_2$O$_3$ for evolution reactions reduce charge transfer kinetics, leading to overall poor PEC performances \cite{tamirat2016using}. Therefore, it is crucial to address these challenges by employing effective enhancement strategies to fully realize the potential of $\alpha$-Fe$_2$O$_3$ in efficient and sustainable photoelectrochemical water splitting. 

Several enhancement strategies, such as structural engineering, surface modification, and doping methods, have been thoroughly studied to address the drawbacks of pristine $\alpha$-Fe$_2$O$_3$ \cite{ali2022innovative,tofanello2020strategies,park2023recent}. Among these strategies, the doping technique is particularly effective in improving the electronic structure, optical properties, and photocatalytic activity \cite{park2023recent}. A variety of mono-dopants, such as N, Sn, Ti, Si, Pt, Zr, Ge, Cr, and Zn, and co-dopants, such as (N, Zn), (N, Ti), and (Pd, Eu, Rb), have been introduced in pristine $\alpha$-Fe$_2$O$_3$ experimentally to improve the PEC performances \cite{ling2011sn,kim2013single,shen2012effect,wang2013ti,morikawa2013photoactivity,annamalai2016sn}. Nevertheless, due to the extensive efforts and complexity of the experimental processes, it remains challenging to identify the most appropriate dopants and understand the mechanisms that enhance optoelectronic properties and photocatalytic activity. To theoretically address these challenges, the first-principles study is considered an essential approach to examine the design of dopants and assess their effects on electronic structures and optical properties \cite{herfeld2021introduction}. For instance, mono-dopants, such as Be, N, Al, Si, Ni, Ti, Zr, Ge, Rh, and Sn, have been studied, along with co-doping combinations, such as (N, Zr), (N, Ti), and (Be, Sn), to enhance the optoelectronic properties and photocatalytic activity of pristine $\alpha$-Fe$_2$O$_3$ for photoelectrochemical water splitting \cite{annamalai2016sn,pan2015ti,kong2015effective,kleiman2010electrodeposited,wu2023hydrogen}. The potential for discovering new dopants and co-dopants in pristine $\alpha$-Fe$_2$O$_3$ is substantial, offering exciting opportunities for significantly improving the PEC performances.

To further explore the enhancement of the PEC performances of pristine $\alpha$-Fe$_2$O$_3$, this work aimed at designing a modified $\alpha$-Fe$_2$O$_3$ incorporating mono-dopants, such as boron (B), yttrium (Y), and niobium (Nb), and co-dopants, such as (B, Y) and (B, Nb), using first-principles density functional theory (DFT). Firstly, we investigated the thermodynamic phase stability of doped $\alpha$-Fe$_2$O$_3$ by computing formation enthalpy ($E_f$) while considering substitutional and interstitial doping methods. The results revealed that the substitutional mono-dopants, Y and Nb, and co-dopants, (B, Y) and (B, Nb), exhibited negative $E_f$ values, suggesting that these dopants are thermodynamically favorable and practically viable for experimental implementation. Secondly, we examined the material characteristics of the modified doped $\alpha$-Fe$_2$O$_3$ by calculating bulk strain ($\mu_B$), microstrain ($\mu_\epsilon$), crystallite size ($D$), and dislocation density ($\delta$) using X-ray diffraction (XRD) patterns and the Williamson-Hall (WH) method. Thirdly, we calculated the electronic band structures, density of states (DOS), and atom ionic charge to analyze the charge carrier localization in the valence band maximum (VBM) and conduction band minimum (CBM), $E_g$, and charge carrier density. The doped $\alpha$-Fe$_2$O$_3$ exhibited a reduced $E_g$, improved charge carrier delocalization, and increased atom ionic charge, demonstrating enhanced free charge carrier density and electrical conductivity. Finally, we determined the optical absorption coefficient ($\alpha$) incorporating the real and imaginary parts of permittivity for both pristine and doped $\alpha$-Fe$_2$O$_3$. The dopants B, (B, Y), and (B, Nb) into $\alpha$-Fe$_2$O$_3$ displayed significant red-shifted light absorption in the visible spectrum. Our results reveal that the proposed dopants and doping methods could substantially enhance the photocatalytic activity of $\alpha$-Fe$_2$O$_3$, making it a promising non-toxic, low-cost, earth-abundant, and durable photoelectrode for efficient photoelectrochemical water splitting.                

%
%

%
%
\section{Computational Methodology} 

This work performed all calculations within the framework of spin-polarized DFT using the self-consistent ab initio method implemented in the Quantum Espresso (QE) code \cite{giannozzi2009quantum}. The exchange-correlation functions were calculated using the general gradient approximation (GGA) and the Perdew-Burke-Ernzerhof (PBE) model \cite{giannozzi2009quantum,giannozzi2017advanced}. The projected augmented wave (PAW) methodology was applied to describe the core electrons, while the valence electrons were defined using Kohn-Sham (KS) single-electron orbitals, which were solved by expanding in a plane-wave basis with a cut-off kinetic energy of 60 Ry \cite{kresse1999ultrasoft}. Marzari-Vanderbilt smearing was used with a value of 0.01 Ry to aid convergence. The Brillouin-zone integration was carried out using the general Monckhorst-Pack grids of 11$\times$11$\times$1, and the DFT-D3 method was employed for Van der Waals force corrections \cite{monkhorst1976special}. To accurately estimate the self-interacting exchange-correlation functions in the d-orbitals of transition metal atoms, the PBE+U framework developed by Dudarev et al.~was utilized in this simulation \cite{dudarev1998electron}. The effective Hubbard correction term, $U_{\rm eff}$, was set to 4.30, as determined symmetrically by Mosey et al.~\cite{mosey2008rotationally}. The optimized lattice parameters and fully relaxed atomic positions of both pristine and doped $\alpha$-Fe$_2$O$_3$ were achieved through variable-cell relaxation calculations using the Broyden-Fletcher-Goldfarb-Shanno (BFGS) algorithm until reaching the atom-acting force convergence of $10^{-3}$ eV\AA$^{-1}$ and the energy convergence of 10$^{-6}$ eV.

%
%
\begin{figure}[H]
    \centering
    \includegraphics[width =0.87\linewidth]{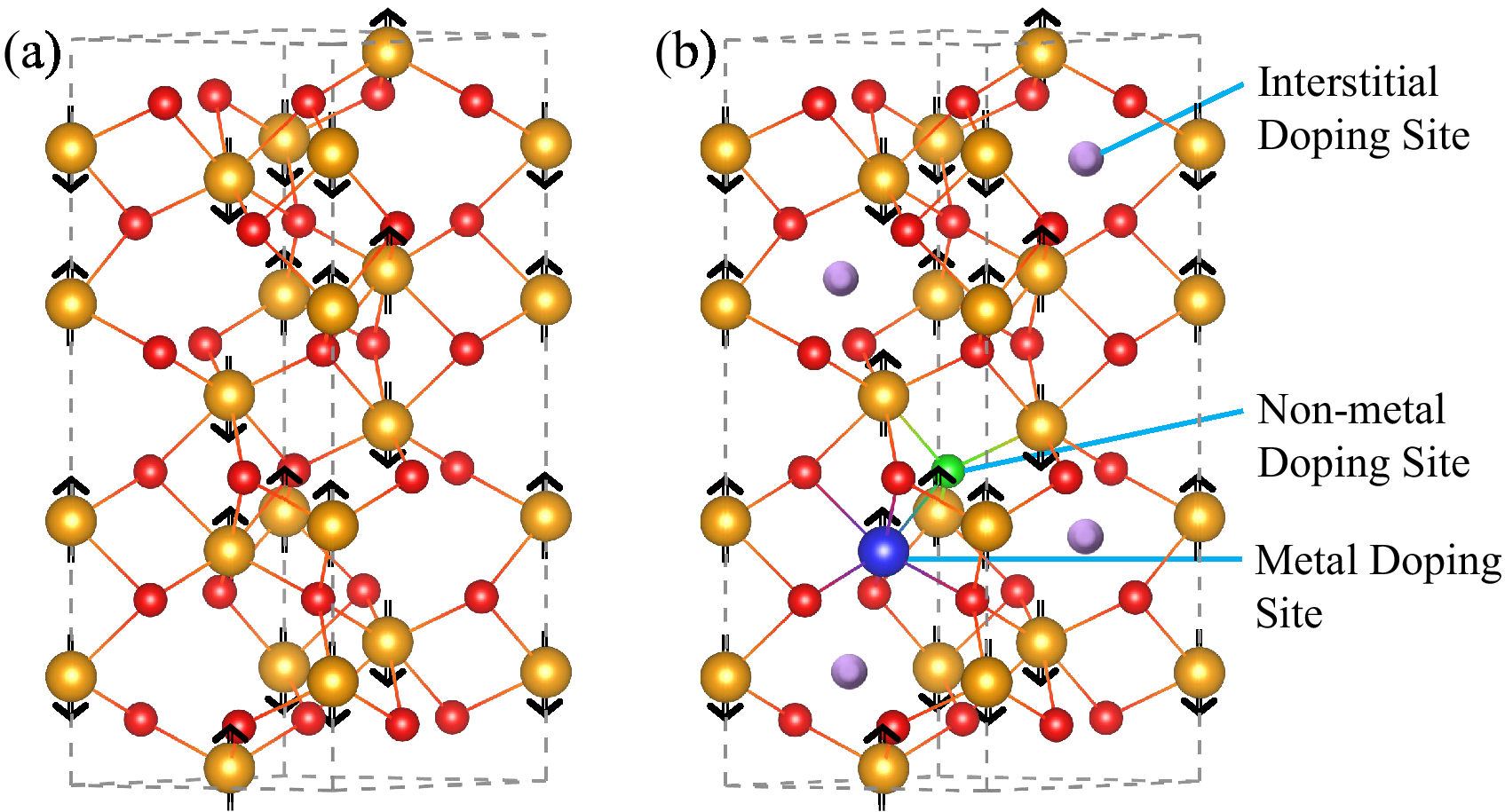}
    \caption{(a) Hexagonal close-packed crystal structure of pristine  $\alpha$-Fe$_2$O$_3$ with the arrow symbol for Fe atom magnetic spin direction at $z$-axis ([0001] plane). The golden and red spheres represent Fe and O atoms, respectively. (b) The substitutional and interstitial doping sites into $\alpha$-Fe$_2$O$_3$ for non-metal (N) and metal (M) dopants. The green, blue, and violet spheres indicate non-metal, metal, and interstitial doping sites, respectively.}
    \label{Fig1}
\end{figure}
%
%

Figure \ref{Fig1}(a) shows the hexagonal close-packed crystal structure of pristine $\alpha$-Fe$_2$O$_3$ (a space group of R$\bar{3}$2/C), consisting of 12 Fe and 18 O atoms. We chose this crystal structure to create doping models for this work. The bulk $\alpha$-Fe$_2$O$_3$ exhibits intriguing weak ferromagnetic properties above the Morin transition temperature at $\sim$263 K and below the N\'eel temperature at $\sim$955 K \cite{smith1916magnetic,tadic2022magnetic}. When the temperature exceeds the N\'eel point, it transitions to a paramagnetic state. However, below the Morin temperature, it possesses a stable antiferromagnetic ground state, characterized by the magnetic spin orientations of the Fe atoms lying in the [0001] plane, as illustrated in Fig.~\ref{Fig1}(a) \cite{tadic2022magnetic,morin1950magnetic}. Substitutional and interstitial doping methods were employed for doping models, as illustrated in Fig.~\ref{Fig1}(b). In the substitutional method, the positions of the up and down spin Fe atoms were considered doping sites for the metals Y and Nb. In addition, non-metal B replaced the positions of O atoms as substitutional doping sites. The unoccupied spaces within the $\alpha$-Fe$_2$O$_3$ crystal structure were designated interstitial doping sites for all dopants, including B, Y, and Nb. To model co-dopants into $\alpha$-Fe$_2$O$_3$, Fe and O atoms were substituted by the metal and non-metal dopants, respectively. 

To investigate the growth mechanism and assess the relative phase stability of doped $\alpha$-Fe$_2$O$_3$, we calculated the impurity $E_f$ for substitutional mono-dopants and co-dopants using the following equations \cite{wang2014band} 
\begin{subequations}
\begin{align}
    &E_{f,{\rm N}}^{\rm sub} = E_{\rm N } - E_{\rm P} - \mu_{\rm N} + \mu_{\rm O},\\
    &E_{f,{\rm M}}^{\rm sub} = E_{\rm M} - E_{\rm P} - \mu_{\rm M} + \mu_{\rm Fe},\\ 
    &E_{f,({\rm N,M})}^{\rm sub} = E_{\rm (N,M)} - E_{\rm P } - \mu_{\rm N} + \mu_{\rm O} - \mu_{\rm M} + \mu_{\rm Fe}. 
\end{align}
\end{subequations}
The impurity $E_f$ for interstitial dopants was determined as 
\begin{equation}
    E_{f,{\rm N/M}}^{\rm int} = E_{\rm N/M} - E_{\rm P} - \mu_{\rm N/M}, 
\end{equation}
where $E_{f,{\rm N}}^{\rm sub}$, $E_{f,{\rm M}}^{\rm sub}$, and $E_{f,({\rm N,M})}^{\rm sub}$ refer to the formation enthalpies of dopants for the substitutional method. Specifically, $E_{f,{\rm N}}^{\rm sub}$ refer to the formation enthalpy for mono-doping with a non-metal (N), $E_{f,{\rm M}}^{\rm sub}$ for mono-doping with a metal (M), and $E_{f,({\rm N,M})}^{\rm sub}$ for co-doping with both a non-metal and a metal (N, M). Additionally, $E_{f,{\rm N/M}}^{\rm int}$ represents the formation enthalpy of non-metal/metal (N/M) dopants when using the interstitial method. The total system energies due to doping elements are denoted as follows: $E_{\rm N}$ for substitute mono-doping with N, $E_{\rm M}$ for substitute mono-doping with M, and $E_{\rm (N,M)}$ for substitute co-doping with both N and M, and $E_{\rm N/M}$ for interstitial doping with N or M dopant. The parameter $E_{\rm P}$ represents the total energy of pristine $\alpha$-Fe$_2$O$_3$ unit cell. The chemical potentials for N, M, O, and Fe are represented by $\mu_{\rm N}$, $\mu_{\rm M}$, $\mu_{\rm O}$, and $\mu_{\rm Fe}$, respectively. It is important to note that the $\mu_{\rm O}$ and $\mu_{\rm Fe}$ values critically depend on the growth mechanism of $\alpha$-Fe$_2$O$_3$, specifically under O-rich and Fe-rich conditions, causing variations in the impurity $E_f$  \cite{kraushofer2018atomic}. To thoroughly examine these variations, we determined $E_f$ for both conditions by calculating $\mu_{\rm O}$ and $\mu_{\rm Fe}$ based on the chemical potential of $\alpha$-Fe$_2$O$_3$ ($\mu_{\rm Fe_2O_3}$) using the following equation
\begin{equation}
    \mu_{\rm Fe_2O_3} = 2\mu_{\rm Fe} + 3\mu_{\rm O}.
\end{equation}
Under the Fe-rich (O-rich) condition, the chemical potential $\mu_{\rm Fe}$ ($\mu_{\rm O}$) is calculated from the energy of a bulk Fe (O) atom, while the corresponding $\mu_{\rm O}$ ($\mu_{\rm Fe}$) is determined using Eq.~(3).

We evaluated several material characterization parameters for pristine and doped $\alpha$-Fe$_2$O$_3$, such as $\mu_B$, $\mu_\epsilon$, $D$, and $\delta$. The value of $\mu_B$ was calculated by the ratio between the optimized unit cell volume ($V_{\rm cell}$) of doped and pristine $\alpha$-Fe$_2$O$_3$. Meanwhile, the X-ray diffraction (XRD) pattern and the Williamson-Hall (WH) method were employed to determine the values of $\mu_\epsilon$ and $D$ \cite{williamson1953x}. To obtain the XRD data for the pristine and doped $\alpha$-Fe$_2$O$_3$, we used Vesta Software considering conventional $\theta$/2$\theta$ scan at the wavelength ($\lambda_{\rm XRD}$) of 1.506 {\AA} (refer to Figs.~S1--S3 for the XRD patterns) \cite{momma2008vesta}. The WH equation is given by \cite{williamson1953x}
\begin{equation}
    \beta{_{hkl}} \cos(\theta) = \frac{K \lambda_{\rm XRD}}{D} + 4 \mu_\epsilon \sin(\theta),
\end{equation}
where $\beta_{hkl}$ is the integral breadth of the peak associated with the ($hkl$) plane in the XRD profiles, $\theta$ is the diffraction Bragg angle, and $K$ is the shape control factor that commonly has a value of $0.9$. The parameter $\delta$ is derived from $D$ using $\delta = 1/D^2$. In order to investigate the interaction of optical photons with electrons in both pristine and doped $\alpha$-Fe$_2$O$_3$, $\alpha$ is calculated from the real $\epsilon_{\rm re}(\omega)$ and imaginary $\epsilon_{\rm im}(\omega)$ parts of permittivity using \cite{modak2014band} 
\begin{equation}
    \alpha(\omega) = 2\sqrt{\omega} \left[\frac{\sqrt{\epsilon_{\rm re}^2(\omega) + \epsilon_{\rm im}^2(\omega)} - \epsilon_{\rm re}(\omega) }{2} \right]^{1/2},
\end{equation}
where $\omega$ is the angular frequency, which can be expressed as $\omega = (2\pi c)/{\lambda}$. Here, $c$ and $\lambda$ are the speed of light and optical wavelength, respectively.

%
%

%
%
\begin{table}[H]
\centering
\caption{Optimized structural lattice parameters and magnetic moment ($\mu_m$) of pristine  $\alpha$-Fe$_2$O$_3$ and doped $\alpha$-Fe$_2$O$_3$ photoelectrodes.}
\resizebox{0.87\textwidth}{!}{%
\begin{tabular}{c|c|c|c|c|c}
\Xhline{3\arrayrulewidth}
    Photoelectrode Name & Method & $a = b$ (\AA) & $c$ (\AA) & $V_{\rm cell}$ ({\AA}$^3$) & $\mu_m$ ({\textmu}B/Fe)\\
    \Xhline{2\arrayrulewidth}
    pristine  $\alpha$-Fe$_2$O$_3$  & PBE & $4.780$ & $13.320$ & $263.374$ & - \\
    pristine  $\alpha$-Fe$_2$O$_3$  & PBE$+$U & $5.147$ & $13.873$ & $318.365$ & 4.168 \\
    B doped $\alpha$-Fe$_2$O$_3$  & PBE$+$U & $5.158$ & $13.954$ & $321.508$ & 4.181 \\
    Y doped $\alpha$-Fe$_2$O$_3$  & PBE$+$U & $5.207$ & $14.021$ & $329.218$ & 4.193 \\
    Nb doped $\alpha$-Fe$_2$O$_3$  & PBE$+$U & $5.214$ & $13.980$ & $329.139$ & 4.124 \\
    (B, Y) doped $\alpha$-Fe$_2$O$_3$  & PBE$+$U & $5.205$ & $14.030$ & $329.177$ & 4.191 \\
    (B, Nb) doped $\alpha$-Fe$_2$O$_3$  & PBE$+$U & $5.210$ & $14.081$ & $331.009$ & 4.190 \\
    \Xhline{3\arrayrulewidth}
\end{tabular}}
\label{Table1}
\end{table}
%
%

%
%
\section{Results and Discussion}
The doped $\alpha$-Fe$_2$O$_3$ requires investigation into the phase stability, optoelectronic properties, and photocatalytic activity. Firstly, we calculated the optimized structure to examine cell parameters, magnetic properties, and the favorable thermodynamic phase through the value of $E_f$. Then, we systematically described the material characterization and electronic structure of pristine and doped $\alpha$-Fe$_2$O$_3$, which aids in understanding the growth mechanism and electronic properties, including $E_g$, charge carrier density, and conductivity. Finally, we presented evidence of improved photocatalytic activity, demonstrating both qualitative and quantitative aspects of this enhancement.   

%
%
\begin{table}[H]
\centering
\caption{Formation enthalpy ($E_f$) of doped $\alpha$-Fe$_2$O$_3$ photoelectrodes incorporating B, Y, and Nb dopants for substitutional and interstitial doping methods.}
\resizebox{0.47\textwidth}{!}{%
\begin{tabular}{c|c|c|c}
\Xhline{3\arrayrulewidth}
    Doping Method & Dopant & \multicolumn{2}{c}{ $E_f$ (eV)} \\
    \cline{3-4}
     & Element & Fe-rich & O-rich \\
    \Xhline{2\arrayrulewidth}
    \multirow{5}{4em}  & B & $3.857$ & $4.611$ \\
      & Y & $-2.657$ & $-3.788$ \\
     Substitutional & Nb & $-3.368$ & $-4.499$ \\
      & (B, Y) & $-1.527$ & $-1.904$ \\
      & (B, Nb) & $-1.991$ & $-2.369$ \\
     \Xhline{2\arrayrulewidth}
     \multirow{3}{4em}  & B & $6.642$ & $6.642$ \\
     Interstitial & Y & $2.899$ & $2.899$ \\
      & Nb & $1.208$ & $1.208$ \\
    \Xhline{3\arrayrulewidth}
\end{tabular}}
\label{Table2}
\end{table}
%
%

\subsection{Optimized Structure and Formation Enthalpy}

We performed a thorough analysis to calculate the lattice parameters of pristine $\alpha$-Fe$_2$O$_3$ using PBE and PBE+U methods to ensure the reliability of our computational approach. For the PBE method, the simulated lattice parameters $a=b$ and $c$ for pristine $\alpha$-Fe$_2$O$_3$ were 4.780 {\AA} and 13.320 \AA, respectively, whereas the experimental values were $a=b=5.04$ {\AA} and $c=13.75$ \AA \cite{pauling1925crystal,finger1980crystal}. The results obtained using the PBE method were underestimated compared to the experimental findings due to excluding the d-orbital self-interacting forces of the transition metal atoms. Using the PBE+U method with $U_{\rm eff}$ set to 4.30, as reported by Meng et al., the computational values were $a=b=5.104$ \AA~and $c=13.907$ {\AA} \cite{meng2016density}. In our simulations using the PBE+U method, the obtained lattice parameters were $a=b=5.147$ {\AA} and $c=13.873$ \AA, and the corresponding magnetic moment ($\mu_m$) was 4.168 {\textmu}$_B$/Fe. The experimental and computational values of $\mu_m$ in the literature are approximately 4.6--4.9 and 4.2 {\textmu}$_B$/Fe \cite{morin1950magnetic,finger1980crystal,meng2016density}. 
Our calculated results closely match with those in the literature, confirming the validity of our methodology to proceed further with the modeling of doped $\alpha$-Fe$_2$O$_3$. Table \ref{Table1} presents all the optimized lattice parameters and $\mu_m$ of pristine and doped $\alpha$-Fe$_2$O$_3$. The value of $\mu_m$ slightly increased for the doped $\alpha$-Fe$_2$O$_3$, except in the case of Nb dopants, where a thermodynamically stable phase occurs in the down-spin position of the Fe atom. Notably, the higher value of $\mu_m$ causes an intense magnetic field based on electron spin polarization and Lorentz force, leading to improved charge carrier separation and reduced non-radiative recombination.   

In order to effectively explore the potential of non-metal and transition metal dopants in pristine $\alpha$-Fe$_2$O$_3$, we strategically chose B as the non-metal dopant and Y and Nb as transition metal dopants for single element doping. We also assessed co-doping combinations of (B, Y) and (B, Nb) to investigate the doping mechanisms involving a trivalent non-metal alongside trivalent and pentavalent transition metal dopants. Table \ref{Table2} shows $E_f$ for all the mono-dopants and co-dopants utilizing substitutional and interstitial doping methods. For the substitutional doping method, the $E_f$ of B dopant was positive under Fe-rich and O-rich conditions, implying its thermodynamic phase instability. In contrast, the mono-dopants, Y and Nb, and the co-dopants, (B, Y) and (B, Nb), exhibited negative $E_f$ values with larger absolute magnitudes under both conditions, demonstrating significant thermodynamic phase stability and suggesting their experimental implementation highly feasible. On the other hand, the interstitial doping method resulted in positive $E_f$ values for B, Y, and Nb dopants, signifying that this approach is endothermic and pristine $\alpha$-Fe$_2$O$_3$ possesses more stability. Thus, it is clear that the interstitial doping method is not a viable option for incorporating B, Y, and Nb dopants into pristine $\alpha$-Fe$_2$O$_3$, emphasizing the importance of favoring substitutional doping strategy for achieving successful experimental outcomes.
                   
%
\begin{figure}[H]
    \centering
    \includegraphics[width =0.91\linewidth]{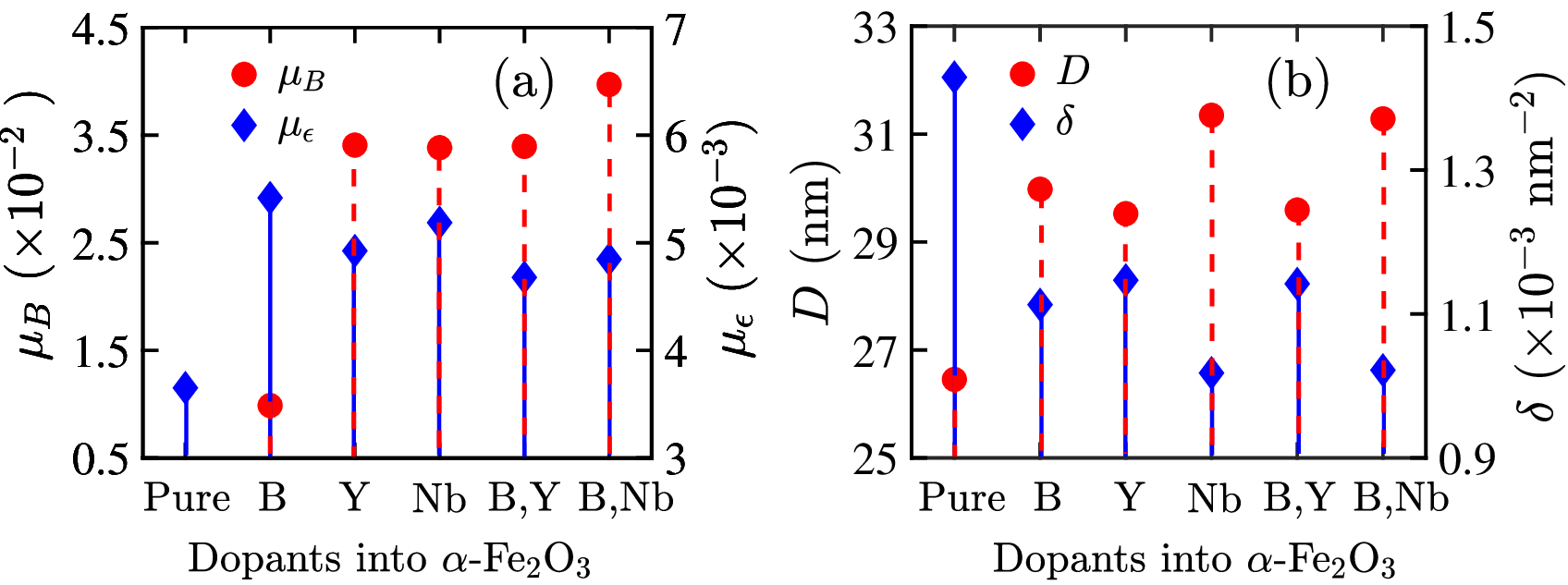}
    \caption{(a) Bulk strain ($\mu_B$) and microstrain ($\mu_\epsilon$), (b) crystallite size ($D$) and dislocation density $\delta$ for pristine and doped $\alpha$-Fe$_2$O$_3$ incorporating B, Y, and Nb dopants.}
    \label{Fig2}
\end{figure}
%
%

\subsection{Material Characterization}

The volume of the unit cell in pristine $\alpha$-Fe$_2$O$_3$ can change when dopants are incorporated. The parameter $\mu_B$ clearly illustrates the changes in volume associated with expansion or compression of doped $\alpha$-Fe$_2$O$_3$, reflecting how the material responds to bulk stress. A smaller value of $\mu_B$ is highly desirable for a photoelectrode, as it signifies a more effective growth mechanism \cite{pun2022effects}. Notably, B dopant possessed the smallest $\mu_B$ value, while co-dopants (B, Nb) had the greatest, as depicted in Fig.~\ref{Fig2}(a). For the mono-dopants, Y and Nb, and the co-dopants (B, Y), the $\mu_B$ value was $\sim$$3.4 \times 10^{-2}$. Remarkably, all dopants exhibited $\mu_B <4$\%, highlighting their potential for practical applications as high-quality $\alpha$-Fe$_2$O$_3$. 

The parameter $\mu_\epsilon$ critically describes the non-uniformity of crystallites and residual lattice strain in materials across different crystal planes, which can significantly influence the optoelectronic and structural properties. A smaller $\mu_\epsilon$ indicates better crystallinity with lower lattice deformation and fewer defects, leading to reduced carrier recombination and improved charge transport efficiency \cite{dhanasekaran2013post}. The experimental value of $\mu_\epsilon$ for pristine $\alpha$-Fe$_2$O$_3$ ranged from $1.51 \times 10^{-3}$ to $6.2 \times 10^{-3}$ \cite{tahir2023investigation,utari2022effect}. By utilizing all the crystal plane peak intensities presented in the XRD pattern shown in Fig.~S1(a), our calculated average $\mu_\epsilon$ was $3.65 \times 10^{-3}$ for pristine $\alpha$-Fe$_2$O$_3$, showing an excellent agreement with the experimental results, as shown in Fig.~\ref{Fig2}(a). The dopants Y, (B, Y), and (B, Nb) consistently displayed smaller $\mu_\epsilon$ values, suggesting that they occupy the most favorable sites and serve as more effective dopants. The Nb dopant, although beneficial, showed a comparatively higher $\mu_\epsilon$, probably due to its surrounding additional under-coordinated O atoms. On the other hand, the B dopant exhibited the highest $\mu_\epsilon$ due to excess ligand bonding relative to the O atom, resulting in stronger interactions with Fe atoms that distort the crystalline lattice. Overall, all dopants demonstrated the values of $\mu_\epsilon$ within the ranges of experimental outcomes for pristine $\alpha$-Fe$_2$O$_3$, indicating their potential effectiveness in structural growth mechanisms and improvements in optoelectronic properties.

The parameter $D$ is critical for measuring the electrical, optical, thermal, magnetic, and chemical properties of the material. A higher value of $D$ leads to increased lattice strain and narrower grain boundary widths, decreasing electrical resistivity and increasing conductivity \cite{iqbal2011study}. Furthermore, the optical properties, such as photoluminescence (PL) intensities and photocatalytic activity, improve with larger $D$. Similar to microstrain analysis, our calculated average value of $D$ for pristine $\alpha$-Fe$_2$O$_3$ was 26.45 nm, which aligns well with the experimental range of 21 to 32 nm, demonstrating the reliability of our simulations \cite{rincon2019evaluation}. Figure \ref{Fig2}(b) illustrates the values of $D$ for all dopants, with the Nb and (B, Nb) dopants exhibiting the largest $D$, leading to significantly improved electrical conductivity and photocatalytic activity. For other dopants, the values of $D$ remained higher than pristine $\alpha$-Fe$_2$O$_3$, eventually improving electrical and optical properties. The parameter $\delta$, inversely proportional to the square of $D$, represents the number of dislocation lines per unit volume within a material that influences various electrical and mechanical properties, including defect sites, charge carrier density, plastic deformation, and tensile strength \cite{oh2019effects}. In Fig.~\ref{Fig2}(b), the $\delta$ values of all dopants were smaller than pristine $\alpha$-Fe$_2$O$_3$, implying a reduced tensile strength and increased plastic deformation. Nevertheless, the smaller $\delta$ resulted in fewer defect sites and larger grain sizes, notably increasing carrier density and decreasing charge transfer resistance ($R_{\rm CT}$), ultimately enhancing PEC performances.

%

\subsection{Electronic Properties}

\begin{figure}[H]
    \centering
    \includegraphics[width =0.91\linewidth]{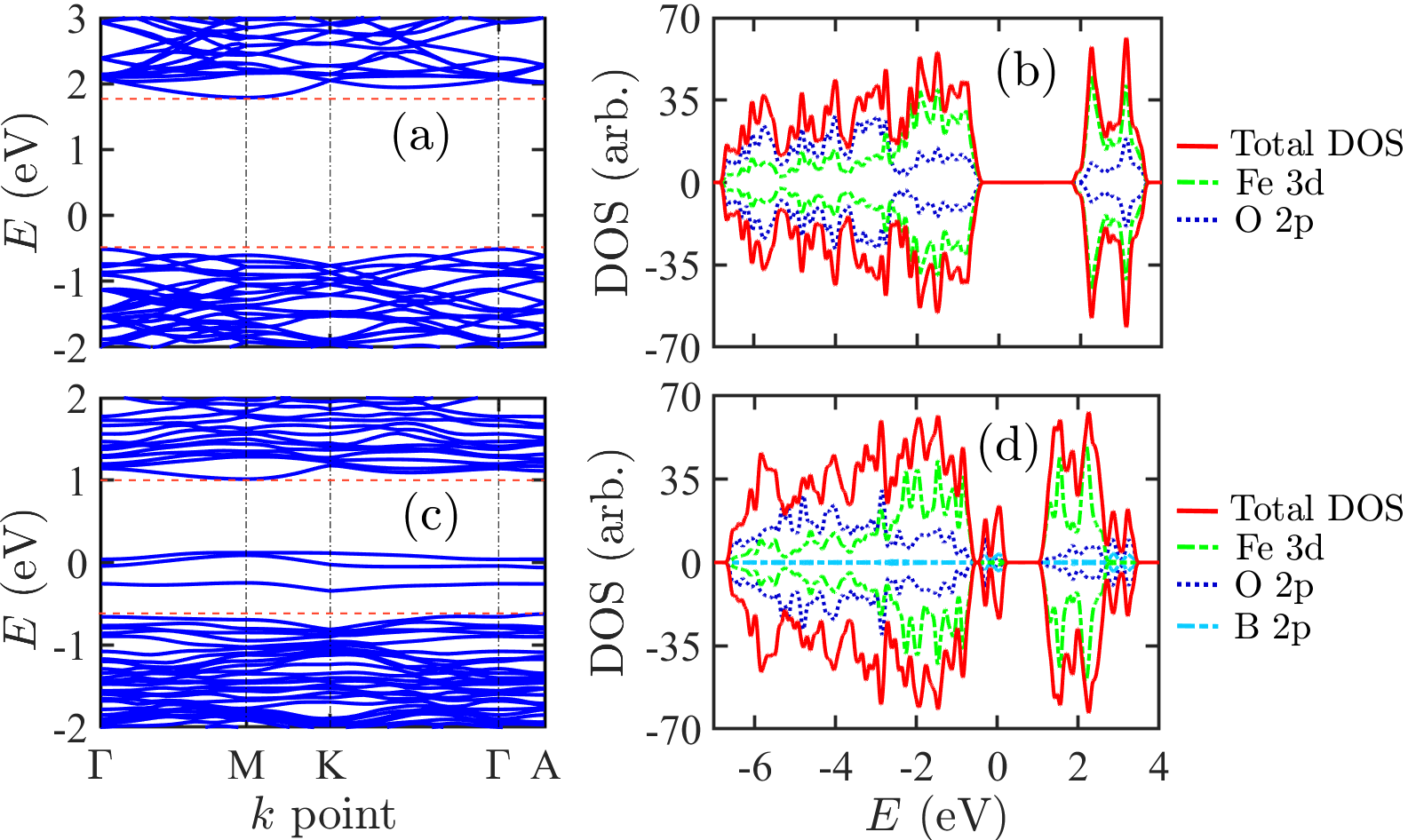}
    \caption{(a) Electronic band structure and (b) density of state (DOS) of pristine  $\alpha$-Fe$_2$O$_3$. (c) Electronic band structure and (d) density of state (DOS) of B doped $\alpha$-Fe$_2$O$_3$. For both of them, the Fermi level is set to zero energy.}
    \label{Fig3}
\end{figure}

We examined the electronic band structure and DOS of pristine $\alpha$-Fe$_2$O$_3$ using both the PBE and PBE+U methods. The PBE method underestimated $E_g$ to be $0.24$ eV, resulting in inaccuracies in the band structure, as shown in Fig.~S4. In contrast, the PBE+U method provided a more accurate estimate of the electronic band structure, reporting an $E_g$ of $2.30$ eV, as illustrated in Fig.~\ref{Fig3}(a). Previous literature has reported an experimental range for $E_g$ between 2.10 eV and 2.40 eV and a calculated value of 2.25 eV, thus assuring the reliability of our approaches and methods \cite{bak2002photo,sivula2011solar,meng2016density,pan2015ti}. Figure \ref{Fig3}(b) presents the DOS of pristine $\alpha$-Fe$_2$O$_3$, highlighting the contributions of the 3d orbitals from Fe and the 2P orbitals from O. The 3d orbitals of Fe atoms primarily contribute to the conduction band (CB), while the valence band (VB) consists of hybridized contributions from the 2p and 3d orbitals of both Fe and O atoms. The near-flatness of the CBM indicates significant localization of charge carriers within the 3d orbitals, leading to a heavy carrier effective mass that complicates electron excitation. Additionally, the relatively large indirect $E_g$  only utilizes a small portion of the solar spectrum, creating substantial challenges for exciting electrons in the 3d orbitals, ultimately leading to very low photoconversion efficiency. 

%
%
%

When the non-metal B dopant replaced the O atom, the $E_g$ reduced to 1.65 eV, significantly enhancing photoabsorption in the visible spectrum by extending wavelengths up to $<752$ nm. In addition, the B dopant introduced acceptor states due to having fewer valence electrons than the O atom, highlighting the potential of B-doped $\alpha$-Fe$_2$O$_3$ as a p-type photoelectrode. However, it is important to note that B doping resulted in several unoccupied impurity bands near the Fermi level, as illustrated in the electronic band structure and DOS in Figs.~\ref{Fig3}(c) and (d). The VB was composed of a combination of B's 2p, O's 2p, and Fe's 3d orbitals, with the O's 2p orbital playing a dominant role. This dominance caused an upward shift of the VB, making it more efficient at generating holes compared to 3d orbitals. Furthermore, the delocalization of charge carriers in the 3d orbitals was advantageous, as the CBM displayed slightly more curvature than that of pristine $\alpha$-Fe$_2$O$_3$. The smaller $E_g$ allowed for the absorption of more photon energy. The impurity band primarily formed from the interaction between B's 2p orbitals and Fe's 3d orbitals, facilitating the transport of charge carriers through this impurity band at relatively low doping concentrations. However, these impurity bands often acted as recombination defects and trap centers for charge carriers, resulting in lower $J_{\rm ph}$ and poor PEC performance. Therefore, it is essential to address and eliminate these impurity bands, and we aim to overcome these challenges using the co-doping technique.     

%
%
\begin{figure}[H]
    \centering
    \includegraphics[width =0.91\linewidth]{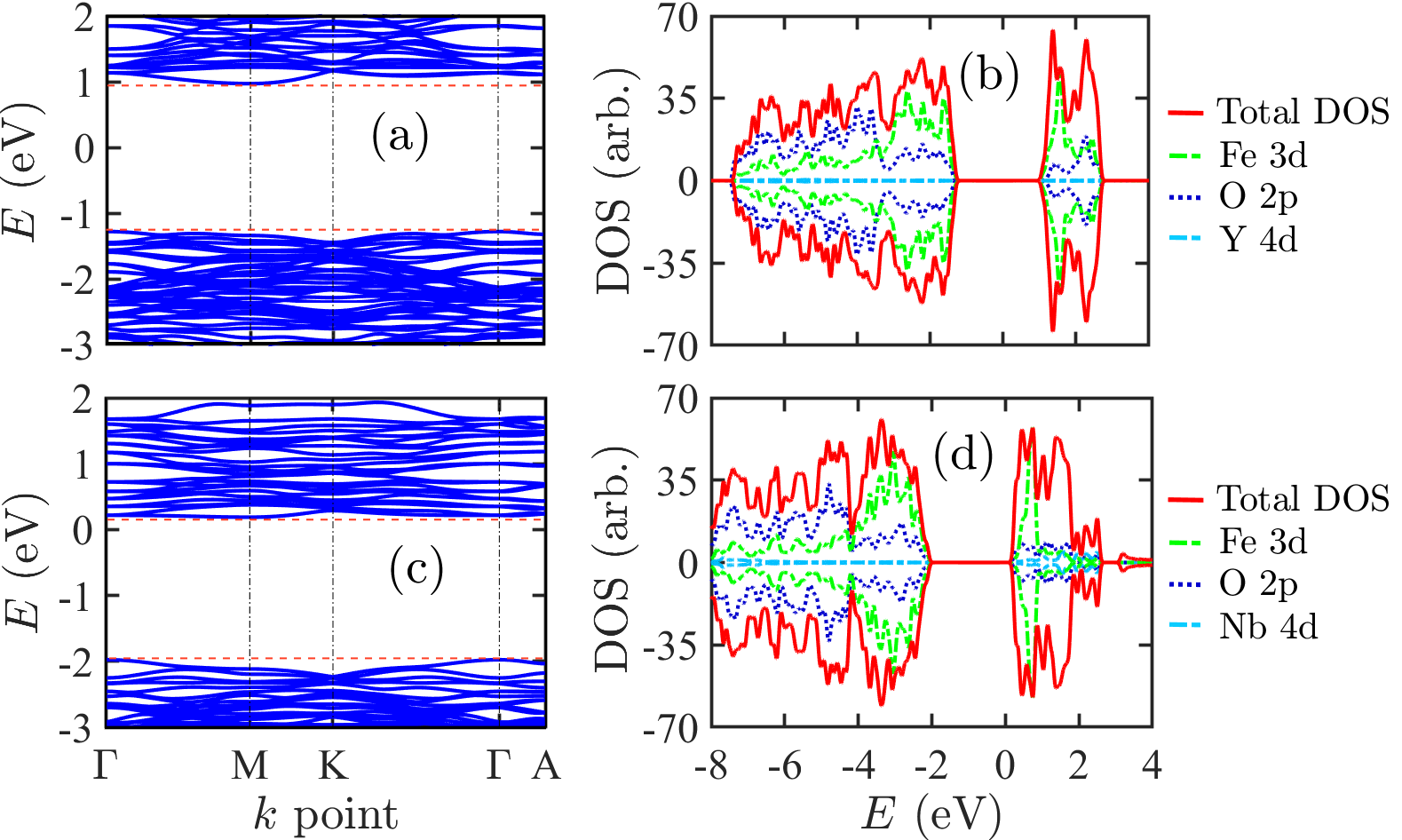}
    \caption{(a) Electronic band structure and (b) density of state (DOS) of Y doped $\alpha$-Fe$_2$O$_3$. (c) Electronic band structure and (d) density of state (DOS) of Nb doped $\alpha$-Fe$_2$O$_3$. For both of them, the Fermi level is set to zero energy.}
    \label{Fig4}
\end{figure}
%
%

Next, we investigated the electronic structure of $\alpha$-Fe$_2$O$_3$ doped with Y and Nb, which acted as donor impurities. Figures \ref{Fig4}(a) and (b) present the electronic band structure and DOS for Y-doped $\alpha$-Fe$_2$O$_3$, respectively. The Fermi level was elevated toward the CB, suggesting its potential use as an n-type photoelectrode. The $E_g$ of Y doped $\alpha$-Fe$_2$O$_3$ slightly decreased to 2.25 eV. However, the presence of the Y dopant significantly increased the charge carrier density and delocalization within the CB, primarily by influencing the 4d orbital of the Y atom. This enhancement improves electrical conductivity and the efficiency of charge transport to the photoelectrode surface. In the case of Nb dopant, Figs.~\ref{Fig4}(c) and (d) depict the electronic band structure and DOS, showing a decrease in $E_g$ by 0.12 eV compared to pristine $\alpha$-Fe$_2$O$_3$. The Fermi level also shifted notably toward the conduction band, further increasing the charge carrier density and reinforcing its role as an n-type photoelectrode. However, the CBM exhibited less curvature than that of pristine $\alpha$-Fe$_2$O$_3$, resulting in a heavier carrier effective mass. This impact adversely affects charge carrier delocalization. Nonetheless, the Nb dopant significantly increased the charge carrier density in the CB, which may help mitigate this adverse effect. Additionally, the VBM with lighter carriers improved charge carrier delocalization, facilitating efficient hole transport toward the photoelectrode surface.

%
\begin{figure}[H]
    \centering
    \includegraphics[width =0.91\linewidth]{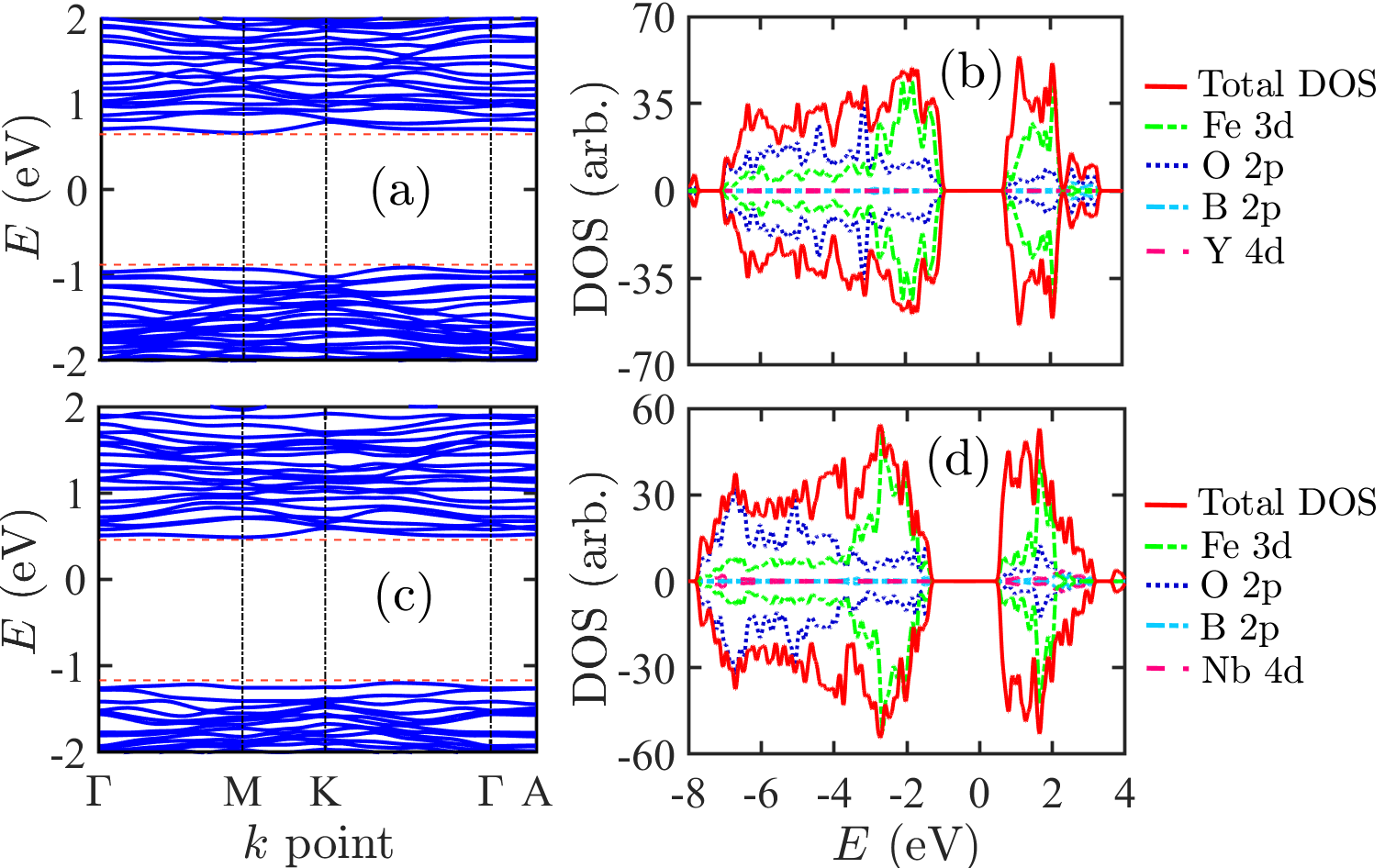}
    \caption{(a) Electronic band structure and (b) density of state (DOS) of (B, Y) co-doped $\alpha$-Fe$_2$O$_3$. (c) Electronic band structure and (d) density of state (DOS) of (B, Nb) co-doped $\alpha$-Fe$_2$O$_3$. For both of them, the Fermi level is set to zero energy.}
    \label{Fig5}
\end{figure}
%
%

Based on previous discussions, incorporating both non-metal and metal dopants into pristine $\alpha$-Fe$_2$O$_3$ exhibited enhanced electronic properties and photocatalytic activity for improved photoelectrochemical water splitting. However, the addition of B as a dopant in $\alpha$-Fe$_2$O$_3$ resulted in the formation of impurity bands, which diminished its effectiveness as a photoelectrode. These challenges can be mitigated by using a combination of n- and p-type impurities as co-dopants. We first investigated the electronic band structure and DOS for (B, Y) co-doped $\alpha$-Fe$_2$O$_3$, as illustrated in Figs.~\ref{Fig5}(a) and (b). In this scenario, impurity bands did not appear between the CB and the VB, and the corresponding $E_g$ reduced to 1.58 eV. This reduction led to extended photoabsorption capabilities in the visible wavelength spectrum, reaching up to $<785$ nm. The donor impurity, Y, played a significant role in (B, Y) co-doped $\alpha$-Fe$_2$O$_3$,  indicating its potential as an n-type photoelectrode. The localization of carriers in the CBM and VBM remained similar to that in pristine $\alpha$-Fe$_2$O$_3$, posing challenges for exciting 3d orbital electrons. Nevertheless, the reduced $E_g$ allowed for adequate solar energy absorption, facilitating the excitation of the 3d orbital electrons for efficient transport within the material. 

Similarly, (B, Nb) co-doped $\alpha$-Fe$_2$O$_3$ was also considered as an n-type photoelectrode. It did not exhibit any impurity bands in the electronic band structure and DOS, as shown in Figs.~\ref{Fig5}(c) and (d), respectively. In this case, the $E_g$ was reduced to 1.69 eV, resulting in excellent photoabsorption and enhanced $J_{\rm ph}$. The CBM of (B, Nb) co-doped $\alpha$-Fe$_2$O$_3$ was more delocalized compared to both the pristine and (B, Y) co-doped $\alpha$-Fe$_2$O$_3$ versions. Furthermore, the Fermi level was positioned closer to the CB, enhancing conductivity and improving the efficiency of charge carrier transport. Overall, introducing (B, Y) and (B, Nb) co-dopants into pristine $\alpha$-Fe$_2$O$_3$ significantly improved its electronic structures and photocatalytic activity, surpassing the effects of mono-doping with B, Y, or Nb alone.                      

%
%
\begin{table}[H]
\centering
\caption{Band gap energy ($E_g$) and electronic charge property of pristine  $\alpha$-Fe$_2$O$_3$ and doped $\alpha$-Fe$_2$O$_3$.}
\resizebox{0.95\textwidth}{!}{%
\begin{tabular}{c|c|c|c|c|c|c|c}
\Xhline{3\arrayrulewidth}
    Photoelectrode Name & \multicolumn{2}{c|}{ $E_g$ (eV)} & \multicolumn{5}{c}{ Bader Charge ($e$C)} \\
    \cline{2-8}
     & PBE & PBE+U & Fe & O & B & Y & Nb \\
    \Xhline{2\arrayrulewidth}
    pristine  $\alpha$-Fe$_2$O$_3$  & $0.24$ & $2.30$ & $+1.822$ & $-1.215$ & - & - & - \\
    B doped $\alpha$-Fe$_2$O$_3$  & $0.05$ & $1.65$ & $+1.727$ & $-1.209$ & $-0.181$ & - & - \\
    Y doped $\alpha$-Fe$_2$O$_3$  & $0.27$ & $2.25$ & $+1.808$ & $-1.226$ & - & $+2.177$ & - \\
    Nb doped $\alpha$-Fe$_2$O$_3$  & $0.14$ & $2.18$ & $+1.748$ & $-1.218$ & - & - & $+2.706$ \\
    (B, Y) doped $\alpha$-Fe$_2$O$_3$  & $0.34$ & $1.58$ & $+1.710$ & $-1.224$ & $-0.185$ & $+2.176$ & - \\
    (B, Nb) doped $\alpha$-Fe$_2$O$_3$  & $0.26$ & $1.69$ & $+1.658$ & $-1.213$ & $-0.187$ & - & $+2.703$ \\
    \Xhline{3\arrayrulewidth}
\end{tabular}}
\label{Table3}
\end{table}
%
%

The electronic charge surrounding each atom provides a quantitative measure to understand the ionic charges created by the atoms in the optimized crystal structure. We conducted a Bader charge analysis on both pristine and doped $\alpha$-Fe$_2$O$_3$ using the framework of DFT with spin polarization enabled. The calculated average ionic charges for the Fe and O atoms were $+1.822e$ and $-1.215e$, respectively. The B atom exhibited a negative charge of $-0.181e$, indicating that the B dopant acted as an acceptor impurity in B-doped $\alpha$-Fe$_2$O$_3$. In the case of mono-doping, Y and Nb atoms contributed ionic charges of $+2.177e$ and $+2.706e$, respectively, to $\alpha$-Fe$_2$O$_3$, resulting in a significant increase in charge carrier density and improved electrical conductivity. Under mono-doping techniques, the ionic charges of Fe slightly decreased, while those of O atoms slightly increased. In contrast, the ionic charge of O in B-doped $\alpha$-Fe$_2$O$_3$ showed negligible reduction. 

Under co-doping conditions, the ionic charges of B, Y, and Nb in $\alpha$-Fe$_2$O$_3$ were nearly the same as those in the mono-doping scenario. However, the ionic charge of the Fe atom decreased by $+0.112e$ and $+0.164e$ for (B, Y) and (B, Nb) co-doped $\alpha$-Fe$_2$O$_3$, respectively. This change is influenced by the 2p orbitals of B and the 4d orbitals of the transition metal dopants. Despite this decrease, the significant donor ionic charges from the transition metal dopants greatly enhanced the overall electronic charge properties, leading to enhanced conductivity and $J_{\rm ph}$ in photoelectrochemical water splitting. Table \ref{Table3} summarizes the electronic charge properties and $E_g$ of both pristine and doped $\alpha$-Fe$_2$O$_3$ considering the PBE and PBE+U methods. The electronic structures of the doped $\alpha$-Fe$_2$O$_3$  under the PBE method are presented in the supplementary material (refer to Figs.~S4 to S9).

%
\begin{figure}[H]
    \centering
    \includegraphics[width =0.93\linewidth]{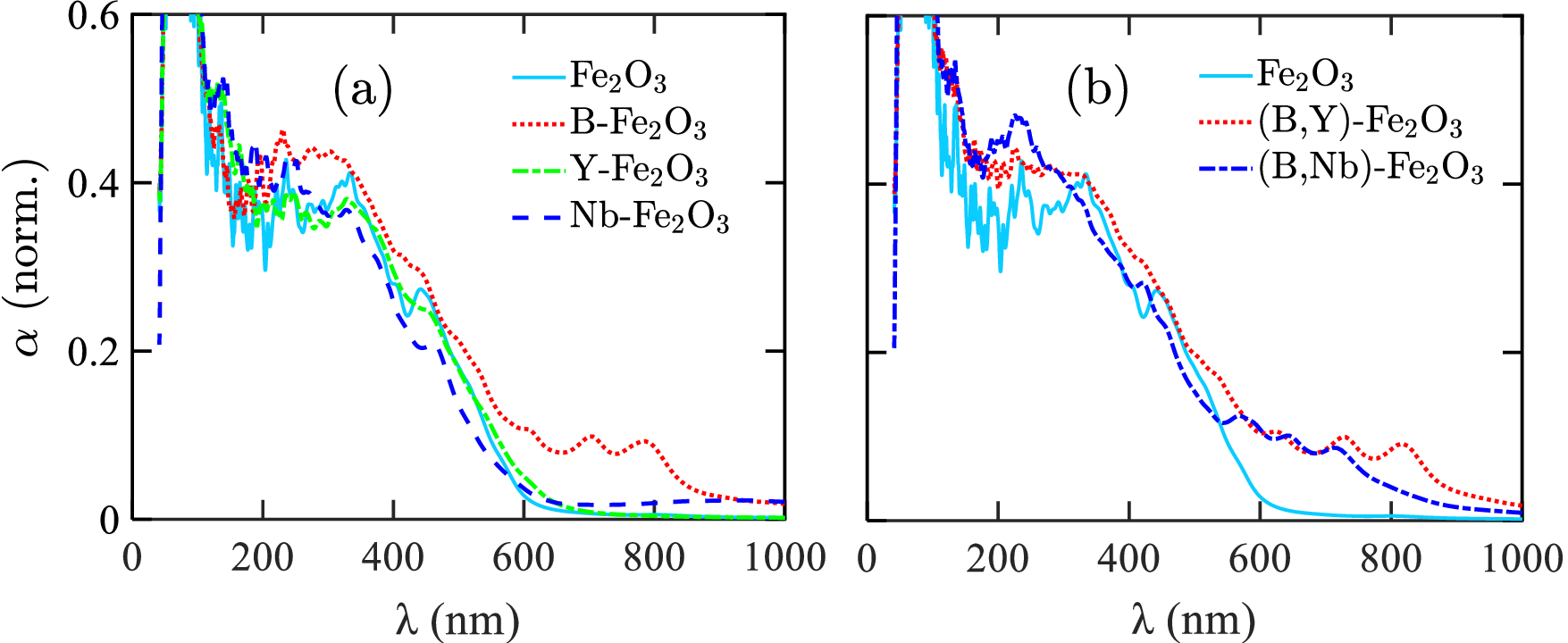}
    \caption{Normalized optical absorption coefficient ($\alpha$) for (a) pristine  $\alpha$-Fe$_2$O$_3$ and mono-doped $\alpha$-Fe$_2$O$_3$ and (b) pristine  $\alpha$-Fe$_2$O$_3$ and co-doped $\alpha$-Fe$_2$O$_3$ photoelectrodes}
    \label{Fig6}
\end{figure}
%
%

\subsection{Optical Properties}

The efficiency of photoabsorption is a crucial factor for photoelectrodes, as it determines how effectively photons interact with electrons in both pristine and doped $\alpha$-Fe$_2$O$_3$ materials. In our analysis of photoabsorption, we calculated the normalized absorption coefficient for both pristine and doped $\alpha$-Fe$_2$O$_3$, as illustrated in Figs.~\ref{Fig6}(a) and (b). Notably, the normalized absorption coefficient of B-doped $\alpha$-Fe$_2$O$_3$ exhibited a significant redshift in the visible spectrum, greatly enhancing its performance compared to pristine $\alpha$-Fe$_2$O$_3$ across the wavelength range of 100 to 800 nm. In contrast, Y-doped $\alpha$-Fe$_2$O$_3$ showed almost the same absorption characteristics similar to those of pristine $\alpha$-Fe$_2$O$_3$, while Nb-doped $\alpha$-Fe$_2$O$_3$ demonstrated improved photoabsorption in the visible spectrum. The most photoactive range for both pristine and mono-doped $\alpha$-Fe$_2$O$_3$ is found between 100 and 390 nm, where excited electrons transit directly between the 2p orbitals of the VB and the 3d orbitals of the CB \cite{gardner1963electrical,shelton2022polaronic}. In the visible spectrum range, the smaller $E_g$ allows for greater sunlight absorption, enhancing photon-electron interactions during the charge carrier transition between the CB and VB, ultimately increasing the absorption coefficient and improving photocatalytic activity.  

The calculated value for absorption coefficient for (B, Y) and (B, Nb) co-doped $\alpha$-Fe$_2$O$_3$ showed significant improvement in the visible spectrum, outperforming pristine $\alpha$-Fe$_2$O$_3$ across the wavelength range of 100 to 1000 nm, as illustrated in Fig.~6(b). In the ultraviolet (UV) region, (B, Nb) co-doped $\alpha$-Fe$_2$O$_3$ exhibited the highest rate of photon-electron interactions for charge carrier transitions between the CB and VB, leading to enhanced photoconversion efficiency. Likewise, (B, Y) co-doped $\alpha$-Fe$_2$O$_3$ displayed higher photoconversion efficiency than pristine $\alpha$-Fe$_2$O$_3$ due to the enhanced delocalization of charge carriers within the CBM and VBM. In addition, the smaller $E_g$ for both combinations of co-dopants played a crucial role in the value of $\alpha$ at longer wavelengths. Therefore, the results reveal that (B, Y) and (B, Nb) are the most effective co-dopants for doping and enhancing the photocatalytic activity of $\alpha$-Fe$_2$O$_3$, making it a promising photoelectrode for efficient and sustainable photoelectrochemical water splitting applications.  

%
%

%
%
\section{Conclusion} 

In summary, we focused on enhancing the electronic and optical properties of $\alpha$-Fe$_2$O$_3$ through mono-doping and co-doping techniques with non-metal elements, B, and metal elements, Y and Nb. By employing the substitutional doping method, we found that the mono-dopants, Y and Nb, and the co-dopants, (B, Y) and (B, Nb), exhibited favorable thermodynamic phase stability and improved material properties. These improvements included a higher $D$ and a preferable $\mu_\epsilon$, enhancing the conductivity and photocatalytic activity. Among the mono-dopants, B-doped $\alpha$-Fe$_2$O$_3$ demonstrated a reduced $E_g$. However, it encountered challenges forming impurity bands between the CB and VB. These challenges were addressed by utilizing co-dopants (B, Y) and (B, Nb) that significantly reduced $E_g$, facilitating light absorption in the visible spectrum. The metal dopants exhibited a higher positive ionic charge, increasing the charge carrier density and $J_{\rm ph}$. In addition, the delocalization of charge carriers in CBM and VBM contributed to the efficient charge transport toward the photoelectrode surface, further improving photocatalytic activity. The absorption of solar energy in the visible region generated more excited charge carriers in the d-orbitals, increasing $J_{\rm ph}$ in photocatalytic water splitting. These findings demonstrate that co-doping with the proposed dopants is the most effective strategy for optimizing growth mechanisms and substantially improving the optoelectronic properties and photocatalytic activity of $\alpha$-Fe$_2$O$_3$, showcasing its potential to advance efficient and sustainable photoelectrochemical water splitting in renewable energy applications.

%
%
\section*{Supplementary material}
\noindent The supplementary material contains the X-ray diffraction patterns, electronic band structures, and density of states of the pristine and doped $\alpha$-Fe$_2$O$_3$ materials.

\section*{Data availability}
\noindent All data in the paper are present in the main text, which will also be available from the corresponding author upon reasonable request.

%
\section*{Author Declaration} 
The authors have no conflicts to disclose.

%
\section*{Acknowledgments} 
The authors gratefully acknowledge financial support from Neural Semiconductor Limited (NSL) and computational support from the Photonics Laboratory, Department of Electrical and Electronic Engineering (EEE), Bangladesh University of Engineering and Technology (BUET).

%
%
\small
\bibliographystyle{ieeetr}
\bibliography{references}

%
\end{document}